\documentstyle[aps,prl,epsfig]{revtex}
\begin{document}
\oddsidemargin--5mm
\large
\newcommand{\bra}{\langle} 
\newcommand{\ket}{\rangle}
\newcommand{\be}{\begin{equation}}
\newcommand{\ee}{\end{equation}} 
\newcommand{\del}{\partial}
\newcommand{\T}{\tau}
\newcommand{\E}{\mbox{erf}}

\leftline{\bf Universal Statistical Behavior of Neural Spike Trains}
\bigskip\bigskip
\leftline{Naama Brenner, Oded Agam, William Bialek}
\leftline{and Rob de Ruyter van Steveninck}
\medskip
\leftline{NEC Research Institute, 4 Independence Way}
\leftline{Princeton, New Jersey 08540}
\bigskip\bigskip
\vfill
\bigskip\bigskip

\hrule

\bigskip\bigskip

\noindent We construct a model that predicts the statistical properties of
spike trains generated by a sensory neuron. The model describes the
combined effects of the neuron's intrinsic properties, the noise in the
surrounding, and the external driving stimulus. We show that
the spike trains exhibit universal statistical behavior over short
times, modulated by a strongly stimulus--dependent behavior over long
times. These predictions are  confirmed in
experiments on H1, a motion--sensitive neuron in the fly visual system.

\vfill\newpage

Neurons in the central nervous system communicate through the
generation of stereotyped pulses, termed action potentials or
spikes \cite{Aidley89,Spikes97}. In many cases these spikes appear to occur as
a random sequence, even under conditions where external sensory stimuli are held
constant. It is tempting to describe these responses in terms
of stochastic models \cite{Holden76}.  On the other hand, neurons in isolation
are described quite accurately by microscopic, deterministic models 
of the form first proposed by Hodgkin and Huxley \cite{HH}. 
How do these microscopic dynamics
relate to the observed statistics of spike trains?  A crucial
ingredient in making this link must be the properties of the
noise that impinges upon the neuron, but we know relatively little about these
noise sources.  Here we argue that, granting certain simple assumptions,
there are some universal statistical behaviors of spike trains
that emerge from a wide class of models, independent of many poorly
known details.
These theoretical results are in good agreement with
data from the fly visual system.

Many isolated neurons generate a 
regular sequence of action potentials when 
a constant current flows across the cell membrane \cite{Jack75,McCormick85}.
This behavior is characterized by the relation between the frequency of
spikes and current, the ``$f/I$ curve.'' 
Microscopic models of neural dynamics, in the spirit of Hodgkin and Huxley, aim
in part at explaining this relation between current and spike frequency
\cite{Troyer97}.
Rather than trying to pass all the way from a microscopic model to a statistical
description, we take the $f/I$ curve as the basic 
phenomenological description of
the cell.  We imagine that input signals $s$ determine a frequency $f(s)$ such
that spikes occur when the time integral of the frequency crosses a
full cycle, as indicated schematically in Fig. 1 \cite{approx}.

\begin{figure}[h]
\vspace{0.2in}
\epsfxsize=12cm
\epsfbox{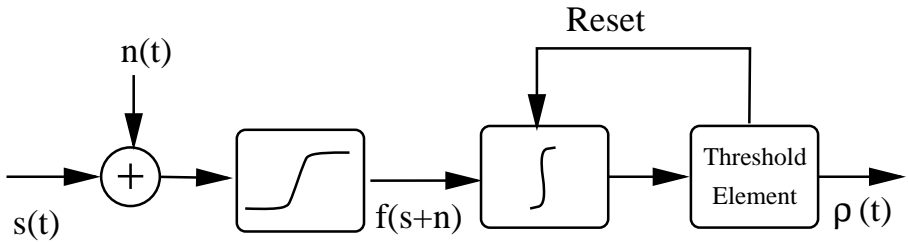}
\vspace{0.4cm}
\caption{
Block diagram of the model for spike generation.
The inputs are a signal
$s(t)$ and a noise $n(t)$, which are added and
passed through the sigmoid-like response
function $f$. 
The modified signal then passes through an integrate-and-fire
element, which generates an action potential when the integrated input
exceeds a threshold (equal to one),
and then resets to zero. The output of this process is a
train of spikes, occuring at the instants
where the threshold was crossed.
}
\label{fig:1}
\end{figure}
\vspace{0.2cm}

Now we would like to ``embed'' this model neuron in a noisy 
environment, such as a complex sensory network. At this step, we
identify $s$ as the external signal to
which the system is sensitive, generally a time dependent
function $s(t)$ \cite{rk0}. The noise is represented by the random function 
$n(t)$, added to the signal at the input of the system.  
The distribution
of this noise defines an ensemble; averages over this ensemble will be
denoted by $\bra \cdots \ket$.
For a constant input $s$, the
firing rate of the neuron $r(s)$ is
\be
r(s) ~=~ \bra f(s+n)\ket. \label{r}
\ee
This relates the cell's $f/I$ curve, which characterizes the
deterministic response to injected current, to the spike rate, which is the {\em
probability} per unit time that the cell will spike in response to the stimulus
$s$.  The relation (\ref{r}) is observable as the average response to repeated
presentations of the same stimulus $s$.  Details of the function $f(s)$
are smoothed by averaging over the noise, and hence are unobservable for a neuron
in its network.
 
In our experiment, a live immobilized fly views various visual stimuli, chosen
to excite the response of the cell H1.  This large neuron is located several
layers back from the eyes, and receives input from many cells.
It is empirically identified as a motion detector, responding optimally to
wide-field rigid horizontal motion, with strong direction selectivity
\cite{Facets89}. 
The fly watches a
screen with a random pattern of
bars, moving horizontally with a velocity $s(t)$. 
We record from H1 extra-cellularly, and the response is
registered as a sequence of spike timings \cite{Ruyter95}.
Figure 2
shows the time-dependent firing rate $r(s(t))$ 
of H1 in response 
to a random signal $s(t)$. This signal is very slow,
and so the relation (\ref{r}) can be
used locally in time.
The plot of firing rate as a function of
the instantaneous signal value (inset of Fig. 2), 
gives us $r(s)$, the noise-smoothed version of $f(s)$.

\begin{figure}[h]
\vspace{0.3in}
\epsfxsize=12cm
\epsfbox{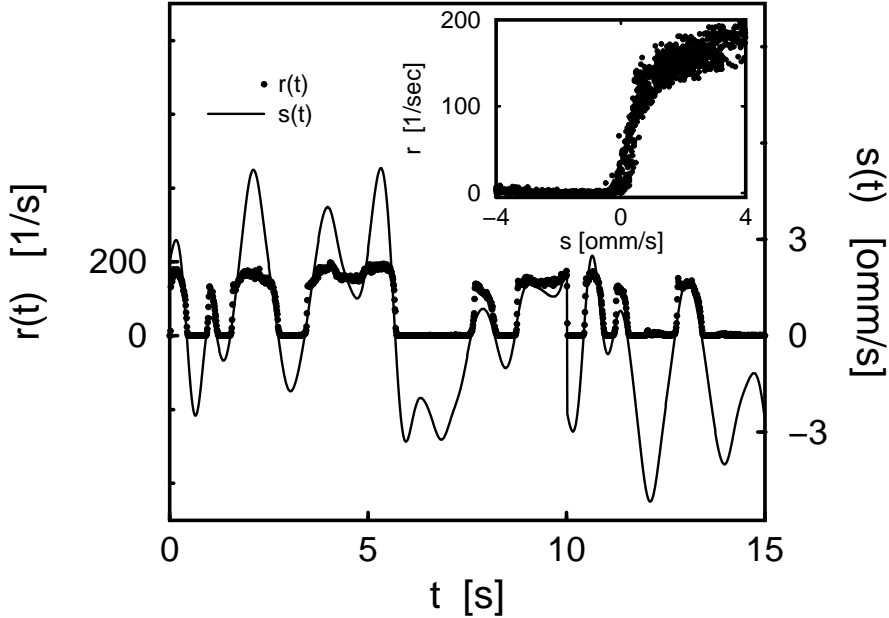}
\vspace{-1.2cm}
\caption{
   Firing rate of H1 as a function of time, averaged over trials:
$r(t)=\bra\rho(t)\ket_{\mbox{trials}}$ [dots],
compared to the input
signal $s(t)$ [solid line].
We repeat the signal many times to obtain a sampling of the noise
ensemble.
In the experiment, a random pattern of vertical bars is moved rigidly
across the visual field with a horizontal angular velocity $s(t)$.
The units of velocity are
spacings of the compound eys lattice (ommatidia) per second.
Inset: instantaneous relation between
$r$ and $s$, which is a noise-smoothed version of the deterministic
response
$f(s)$
(see Eq.\ref{r}).
}
\vspace{-0.2cm}
\label{fig:2}
\end{figure}

\bigskip
\bigskip

Even for the simple model of Fig. 1, the statistical structure of spike trains
can be complicated, dependent upon details in the statistics of the noise
$n(t)$ and the precise form of the function $f(s)$.  Universal behavior
emerges, however, if we assume that the 
stationary noise $n(t)$ is 
characterized by a correlation time, $\xi_n$,
much shorter than the typical inter-spike interval, $1/r$ \cite{rk1}. 
With this assumption, all
the formulas presented below 
follow from the
model via straightforward calculations.
We consider first the case of a constant input signal,
$s(t)\!\equiv\!s$, so that the firing rate  is
constant,  $r(s)\!\equiv \!r$.

Figure 3 shows the theoretical results together 
with the experimental data
from the fly, for three statistical characteristics: the interval
distribution (3a), the autocorrelation function (3b), and the
number variance (3c). Data are presented in gray, and theory in solid black
lines. All three characteristics depend on two parameters, one of which is
the average firing rate $r$. 
It is convenient to
rescale time to dimensionless units $\tau\!=\!r t$. In these units, our
theory has one dimensionless parameter:
\be
\gamma(s) ~=~\xi_n \frac{\bra \delta f(s+n)^2\ket}{\bra f(s+n)\ket},
\ee
\noindent where $\bra \delta f(s\!+\!n)^2\ket\!=\!\bra f(s\!+\!n)^2\ket \!
-\! \bra
f(s\!+\!n)\ket ^2$.
This parameter depends on the correlation time of the noise, $\xi_n$,
and on
the first two moments of $f(s\!+\!n)$ averaged over the noise.
We will see that $\gamma$ describes the variability of the
neural response, and can interpolate between a highly regular response
(small $\gamma$) and a Poisson--like behavior (large $\gamma$). 

\bigskip
\bigskip

\begin{figure}[h]
\vspace{0.3in}
\epsfxsize=16cm
\vspace*{-0.6cm}
\epsfbox{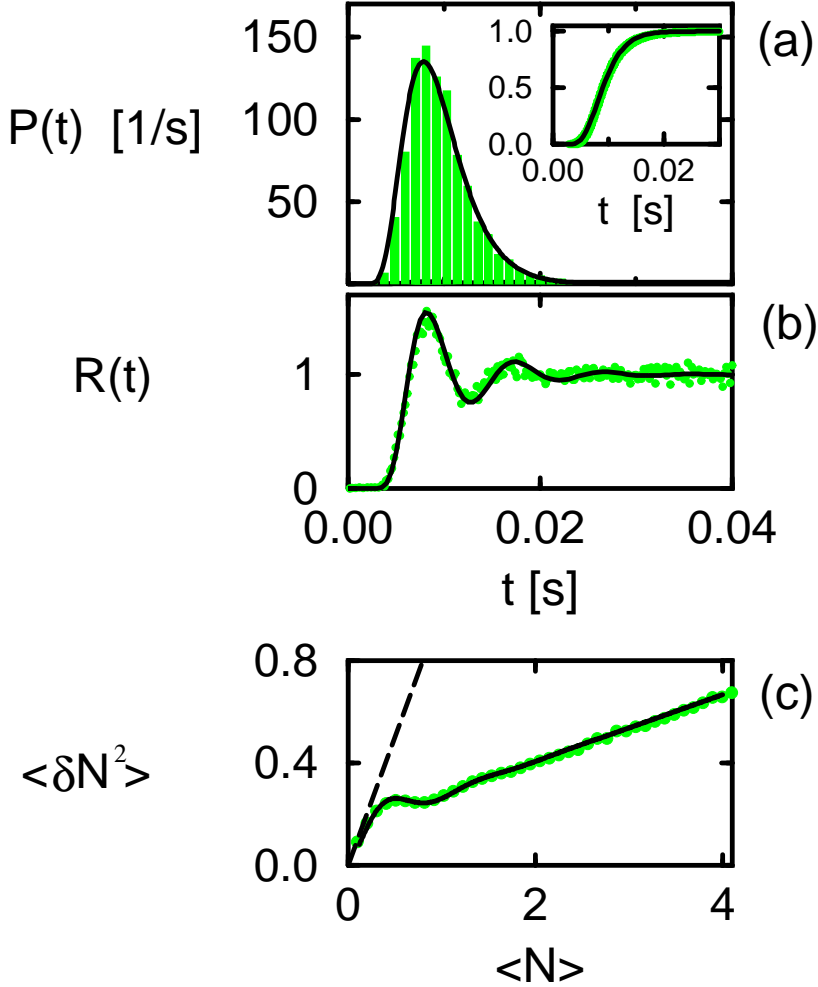}
\caption{
Statistical properties of spike trains in the universal regime:
experiment and theory. All data are taken from an experiment in which
dark
and light bars forming a random pattern, move across the screen with a
constant velocity of 0.12 omm/s.
(a)
Probability distribution of the intervals between successive spikes.
Histogram of intervals from the experiment [gray bars], and
Eq. \ref{ps} [line]. Inset: fraction of intervals of length less than
$t$ [gray line], and integral of Eq. \ref{ps} [line]. (This way of
presenting the data on intervals involves no binning.)
(b) Autocorrelation function calculated from the data
[gray dots], and Eq. \ref{R} [line].
(c) Number variance as a function of number mean, as calculated from the
data [gray dots], and Eq.\ref{num_var} [line].
For comparison, the number variance of a Poisson process, a straight
line of slope 1, is shown by a
dashed line.
}
\label{fig:3}
\end{figure}

\newpage

The
distribution of intervals between neighboring spikes is (Fig. 3a)
\cite{Johannesma68}:
\be
P(\tau) ~~=~~ \frac{\left(1+1/\tau \right)}
{\sqrt{8 \pi \gamma \tau}}
 ~~{\Huge e}^{-\frac{(\tau-1)^2}{2\gamma\tau}}. \label{ps}
\ee
For very short intervals, $\tau\!\ll\! 1$, the distribution decays
strongly: $P(\tau)\sim e^{-1/2\gamma\tau}$; this is  
related to neural refractoriness, but it is not a simple absolute refractory
period. For large intervals,
$P(\tau)\sim e^{-\tau/2\gamma}$. 
The width of the
distribution relative to its mean can be quantified by the coefficient of
variation 
(the standard deviation divided by the mean \cite{Hagiwara54}); 
for our distribution,
this quantity is governed by $\gamma$, 
varying from zero to a constant larger than one as $\gamma$
increases. For the data presented here, $\gamma\approx 0.1$.

We can think of the spike train as a sequence of pulses at times $t_i$,
$\rho(t) = \sum_i \delta (t-t_i).$
Then the (dimensionless)
autocorrelation function of  the spike train, 
$R(\tau)=\bra\rho(0)\rho(\tau)\ket/\bra\rho(0)\ket^2$, is  (Fig. 3b):
\be
R(\tau)  ~~=~~ \sum_{k\neq 0} 
\frac{\left(1+k/\tau \right)}
{\sqrt{8 \pi \gamma \tau}}
 ~~{\Huge e}^{-\frac{(\tau-k)^2}{2\gamma\tau}}  ~~~~~~~~~~~\tau > 0, \label{R}
\ee
with a symmetric expression for $\tau < 0$.
The correlation function is composed
of an infinite sum of functions similar to the
interval distribution, with shifted peaks.
The number of peaks that can be resolved  is proportional to $1/\gamma$:
for small $\gamma$ there is a pronounced 
``ringing'' in the correlation function,
as expected for a regular spike train, while for large $\gamma$ this structure is
absent, indicating an irregular or nearly Poisson sequence of spike times.

The number of spikes in a time interval of length $\tau$ is
$N(\tau)=\int_0^{\tau} \rho(\tau') d\tau'$. 
It is a random variable and its variance
can be expressed as
an integral of the correlation function: 
$\bra \delta N^2\ket =
2\int_0^{\tau} (\tau-\tau')R(\tau')d\tau'$ \cite{Spikes97}. 
Using Eq. (\ref{R}),
we find (Fig. 3c):
\begin{eqnarray}
\bra \delta N
^2(\tau)\ket ~=~  \gamma \tau + 
\frac{1}{2\pi^2}
\sum_{m\neq 0} \frac{1-e^{i2\pi m\tau-2\pi^2 m^2\gamma\tau}}{m^2(1\!+\!
i\pi \gamma m )}. \label{num_var}
\end{eqnarray}
The number variance
consists of a linear term, a constant,
and an infinite sum of oscillating terms.
The linear term
comes from the integration of the $\delta$-function part
of the correlation function, at $\tau\!=\!0$ (omitted from Eq. \ref{R}). 
The oscillatory term decays exponentially, therefore for
long times $\bra\delta
N^2\ket \sim \gamma\tau+c/2$, with 
$c=1/3+\gamma^2-\gamma\coth({1/\gamma})$.

Imagine now that the input signal has a time dependence, $s(t)$, and consider
first the case where it is very slow. Then, the description given above
holds over short time scales.  
Figure 4a shows the autocorrelation function of
spike trains from four different experiments, with different input 
signals. Scaling the different data plots to dimensionless time
$\tau\!=\!r t$, we see that they overlap over short times 
($\tau<2$): the short--time behavior is {\it universal}.
This universality cannot be explained by 
the existence of a constant refractory period:  the
plots are in dimensionless time units, and the region of universality extends
well beyond any reasonable absolute refractory time.  Quantitative universality
requires, moreover, that the value of $\gamma$ be similar for these very
different stimulus conditions.  It seems likely that this approximate constancy
of $\gamma$ occurs by adaptation of the fly's visual system to the
different stimulus ensembles.

\begin{figure}[h]
\epsfxsize=15cm
\vspace{-1.2cm}
\epsfbox{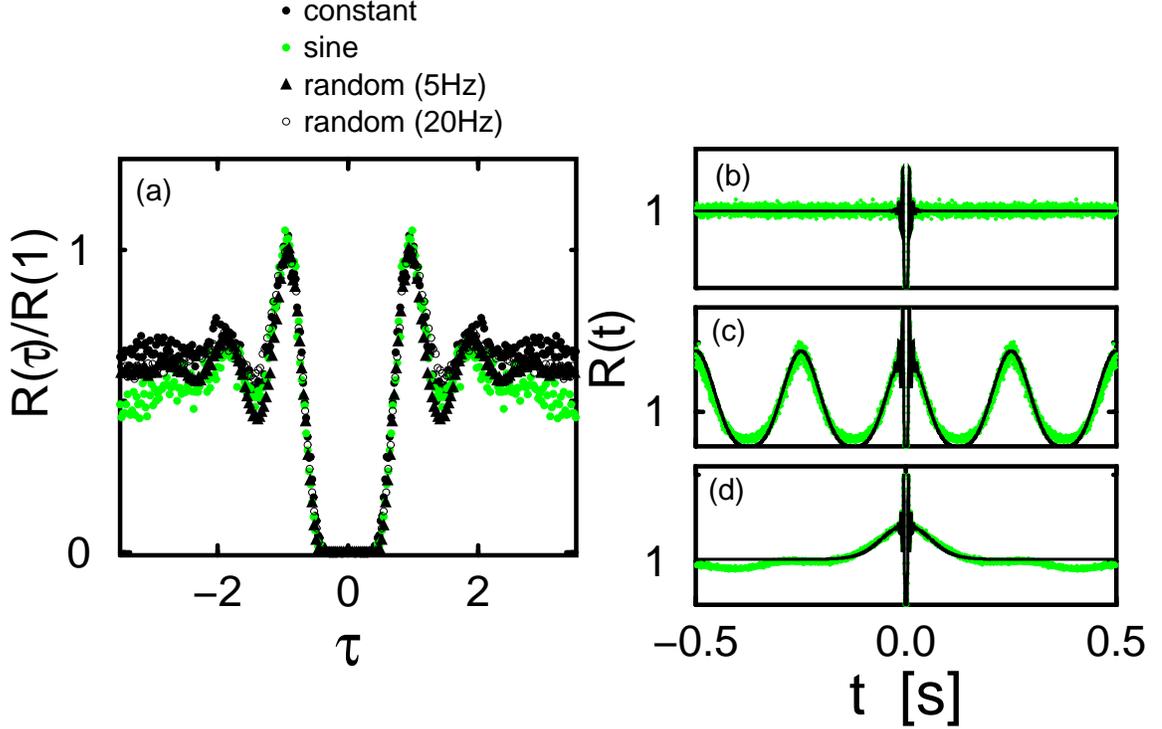}
\caption{
  Correlation functions in the universal (a) and non-universal (b-d)
regimes.
(a) Data from 4 different experiments are superposed. The time axis is
normalized to dimensionless units, and the
horizontal axis is
normalized by the value at the first peak.
The graphs are very similar for
short times, indicating the short--time universality, and begin to
depart at longer times. (b-d)  
Experimental data 
[gray dots] are compared to 
theoretical results [black lines] for the different signals:
constant signal (b), 
Sine wave signal of period 0.25 sec (c),
random
signal with a 5Hz bandwidth (d). 
Theoretical curves
in (b-d) are calculated from known properties of the input signal
by Eq. \ref{RC} with no adjustable parameters.
}
\label{fig:4}
\end{figure}

\bigskip
\bigskip

On longer times, as 
the oscillations of the universal regime decay, 
the curves begin to  reflect correlations in the input
signal (Fig. 4b-d).  This behavior can be predicted from the model
to take the approximate form \cite{rk3}
\be
R(\tau) ~\approx~ 
\overline{r(s(0))r(s(\tau))}
~\sum_{k\neq 0} \frac{(1+k/\tau)}{\sqrt{8\pi \gamma\tau}}~
e^{-\frac{(\tau-k)^2}{2\gamma\tau}}.~~ 
\label{RC}
\ee
Note that the short--time behavior (in the sum) involves only the parameter
$\gamma$ that can be fit in the universal regime, while the correlation function
of the rates can be computed from the input/output relation in the inset of Fig.
2. In many cases simple approximation to this relation (e.g., a step
function) give accurate results for the correlation function;  thus we can
predict correlation functions, at least approximately, with no new parameters.
Figure 4 shows the long--time 
correlation function for various input signals:
a constant signal (4b), 
a sine wave of period 0.25 sec (4c),
and a random signal of bandwidth 5 Hz (4d).
The differences between the various signals show up clearly in the
long-time behavior of $R(t)$, and follow the
theoretical prediction of Eq. \ref{RC} (black lines).  


In conclusion, we have constructed a simple model that
predicts the
statistical behavior of spike trains in the neuron H1 of the fly visual
system. The model is insensitive to the microscopic details
of the spiking neuron; these are represented phenomenologically by the
frequency-current response $f(s)$. When the model neuron is embedded 
in a noisy environment, the spike trains exhibit
universal statistical behavior over short time scales, in which only the
first two moments of $f(s)$ are important. These are represented in the
theory by the two parameters
$r$ and $\gamma$.
In any particular case,  it remains a challenge to understand the microscopic
origins of these parameters, but clearly many different microscopic
models can generate the same value of $\gamma$, 
and hence the same statistics for
the spike train as measured by the inter--spike interval distribution and the
short--time behavior of the autocorrelation function.  On longer time scales, this
universal behavior is modulated by  the statistical properties of the input
signal. The independence of the model on details and the weakness of the
assumptions made, suggest that the results may be valid for a wider class of 
neurons.


\begin{thebibliography}{99}
\bibitem{rk0} The signals used in this work were relatively slow, so
that the input $s(t)$ can be directly identified with the motion on the screen.
If this motion is very fast, filtering mechanisms prior to the
arrival of the signal at H1 become important, and the effective input
signal should be identified as the filtered one.
\bibitem{rk1} The spike train is mathematically defined as
$\rho(t) = \sum_k~ \delta\left(\int_0^t f[s(u)+n(u)]du-k\right) f[s(t)]$. 
Our assumption about the noise having short--range
correlations implies that
integrals such as  $\int^t f[s+n(u)]du$ are approximately 
Gaussian, by the central limit theorem.
\bibitem{rk3} Time may be discretized to steps of size $\xi_n$, by which
our problem becomes analogous to that of a spin chain having real
values. The calculation of statistical properties is then equivalent to
the calculation of
expectation values over all spin configurations, with an appropriate
Hamiltonian. For a constant $s$, the Hamiltonian is that of independent
spins, whereas a time--varying signal $s(t)$ 
induces correlations among spins in the effective Hamiltonian.
In some cases, methods of statistical mechanics (i.e. renormalization
techniques) can be used to solve the problem. A
universal behavior and its decoupling from a signal-dependent part
appears when 
the noise is strong enough, and varies fast enough compared to the input
signal.
\end{thebibliography}
\end{document}